# Quantum well confinement and competitive radiative pathways in the luminescence of black phosphorus layers

Authors

Etienne Carré[1,2], Lorenzo Sponza[1], Alain Lusson[2], Ingrid Stenger[2], Sébastien Roux[1,2], Victor Zatko[3], Bruno Dlubak[3], Pierre Seneor[3], Etienne Gaufrès[4], Annick Loiseau[1] and Julien Barjon[2]*

[1] Laboratoire d'Étude des Microstructures (LEM), UMR 104 CNRS-Onera, Université Paris Saclay, 92322 Châtillon, France.
[2] Groupe d'Etude de la Matière Condensée (GEMaC), CNRS-UVSQ, Université Paris Saclay, 45 Avenue des Etats-Unis 78035 Versailles, France
[3] Unité Mixte de Physique, CNRS -Thales, Université Paris-Saclay, 91767 Palaiseau, France.
[4] Laboratoire Photonique Numérique et Nanosciences, CNRS - Institut d'Optique - Univ. Bordeaux, 33400 Talence, France

*Correspondance : julien.barjon@uvsq.fr

**Abstract:**

**Black phosphorus (BP) stands out from other 2D materials by the wide amplitude of the band-gap energy ($\Delta E_g$) that sweeps an optical window from Visible (VIS) to Infrared (IR) wavelengths, depending on the layer thickness. This singularity made the optical and excitonic properties of BP difficult to map. Specifically, the literature lacks in presenting experimental and theoretical data on the optical properties of BP on an extended thickness range. Here we report the study of an ensemble of photoluminescence spectra from 79 passivated BP flakes recorded at 4 K with thicknesses ranging from 4 nm to 700 nm, obtained by mechanical exfoliation. We observe that the exfoliation steps induce additional defects states that compete the radiative recombination from bound excitons observed in the crystal. We also show that the evolution of the photoluminescence energy *versus* thickness follows a quantum well confinement model appreciable from a thickness predicted and probed at 25 nm. The BP slabs placed in different 2D heterostructures show that the emission energy is not significantly modulated by the dielectric environment.**

## Introduction

Confinement effects in 2D materials have attracted considerable interest because of their layered and exfoliable structures that allow angstrom scale increment of the thickness, from monolayer to bulk. Hence, ultrathin layers of 2D semiconductors, such as the $MoS_2$ and other

TDMCs, have demonstrated stunning physical properties, of high significance for optoelectronics and spintronics[1,2]. In this context, black phosphorus (BP), firstly exfoliated in 2014[3,4], stands out from other 2D materials by the peculiar and strong confinement effects on its physical properties. One emblematic feature is the wide amplitude of the band-gap energy ($\Delta E_g$) that sweeps an optical window from Visible (VIS) to Infrared (IR) wavelengths, depending on the number of layers. Indeed, the optical activities in bulk and monolayer BP are centered at 0.3 eV and 2 eV[5], respectively, corresponding to an amplitude energy *$\Delta E_{g-BP}$ = 1.7 eV* considerably wider compared with that observed in the MoS$_2$ family, roughly *$\Delta E_{g-MoS2}$* = 0.3 eV[6]. Moreover, this variation of *$\Delta E_{g-BP}$* spreads over a large layer number and, for instance, a significant photoluminescence (PL) energy modulation is still observed in the range of 20-30 BP layers.[7]

These singularities make the optical and excitonic properties of BP difficult to map. Specifically, the literature lacks in presenting experimental and theoretical data on the optical properties of BP on an extended thickness range. The difficulty to probe BP samples is, furthermore, exacerbated by the poor sensitivity of the detectors in the IR optical range and by the fast and intrinsic photodegradation of BP when placed in ambient conditions.[8,9] The intercrossing of dielectric, geometrical, defects and mechanical effects with quantum confinement effects also need to be clarified in the ultrathin, thin and bulk-like exfoliated BP flakes. Theoretical[5,10] and experimental[11–15] studies have been focusing on the behavior of few layers BP emitting in the visible / NIR range. In particular, few works analyzed the strong and anisotropic absorption/emission bands associated to the direct band gap in BP in the 1-4 layers regime[11,12,16]. BP samples were also investigated in the semi-bulk and bulk-like thickness ranges, i.e. arbitrary 4 nm to 40 nm and over 400 nm.[17,18] The study of infrared luminescence of thicker layers of BP has accelerated, partly driven by the emergence of infrared optical applications such as BP based photodetectors and lasing.[7,17,19,20] More recently, the fingerprints of bulk BP PL spectra at 2K were also reported, showing the existence of free and bound excitons energy around a refined value of the bandgap of 0.287 eV at cryogenic temperature[21].

Here we report on an unprecedented ensemble of low temperature PL spectra from 79 passivated BP flakes with thicknesses ranging from 4 nm to 700 nm, obtained by mechanical exfoliation. We observed that the exfoliation steps induce additional defects states that compete the radiative recombination from bound exciton observed in the as-grown crystals. We described our data set with a quantum well model applied to a thickness dependent BP slab and we identified a correlation between the optical band gap energy variation and the footprint of the electron-hole wave function within a BP bulk crystal. Finally, we evidenced that the dielectric environment does not modulate the photoluminescence energy from BP layers sandwiched in different 2D heterostructures.

**Main text**

**Defect-related emission from exfoliated crystals.**

The BP crystals were purchased from HQ Graphene and were mechanically exfoliated by an usual repeatable stamping process in a glovebox under argon atmosphere (<0.5 ppm $O_2$, <1 ppm $H_2O$) to prevent BP degradation. The thickness of the flakes was characterized by combining optical and Atomic Force Microscopy (AFM), both installed in the glovebox. A custom-built transfer box was used to transfer the samples into an Atomic Layer Deposition (ALD) chamber to perform a homogeneous deposition of a 10 nm $Al_2O_3$ passivation layer. Further details on the sample preparation and PL characterization can be found in the supplementary information file [SupMat1]).

In Figure 1 we compare representative photoluminescence spectra recorded at 4K from: an as-received BP crystal (1a); a crystal resulting from a single-stamping exfoliation deposited on a Si/SiO$_2$ substrate (1b); and a thick BP flake (700 nm) obtained from a multiple-stamping exfoliation deposited on substrate (1c). The corresponding photography and optical images are shown in Figure 1d. As described in[21], the PL of the as-received crystal consists of two structures. A fine emission signal is formed mainly of free and bound excitons (I°X) lines, and is highlighted by the pale-blue strip in the Figure 1a-c. A second contribution at about 20 meV lower energy is related to defect-assisted recombination. Because size effects are negligible at this thickness, as we will demonstrate below, the PL fingerprints are expected to look similar in all three samples. Surprisingly, even if the overall PL emission bands fall in the same tight energy interval (50 meV), the line shape is significantly modified. The excitonic fine structure of BP crystal (sharp peak) clearly disappeared leaving a broadened emission line for the sample resulting from multiple-stamping exfoliation (Figure 1c). The characteristics of this emission peak (line widths ranging from 15 meV to 45 meV) are consistent with the PL spectra reported in the literature for different thick flakes obtained via exfoliation[7,17,19,20], but its origin has not been elucidated yet.

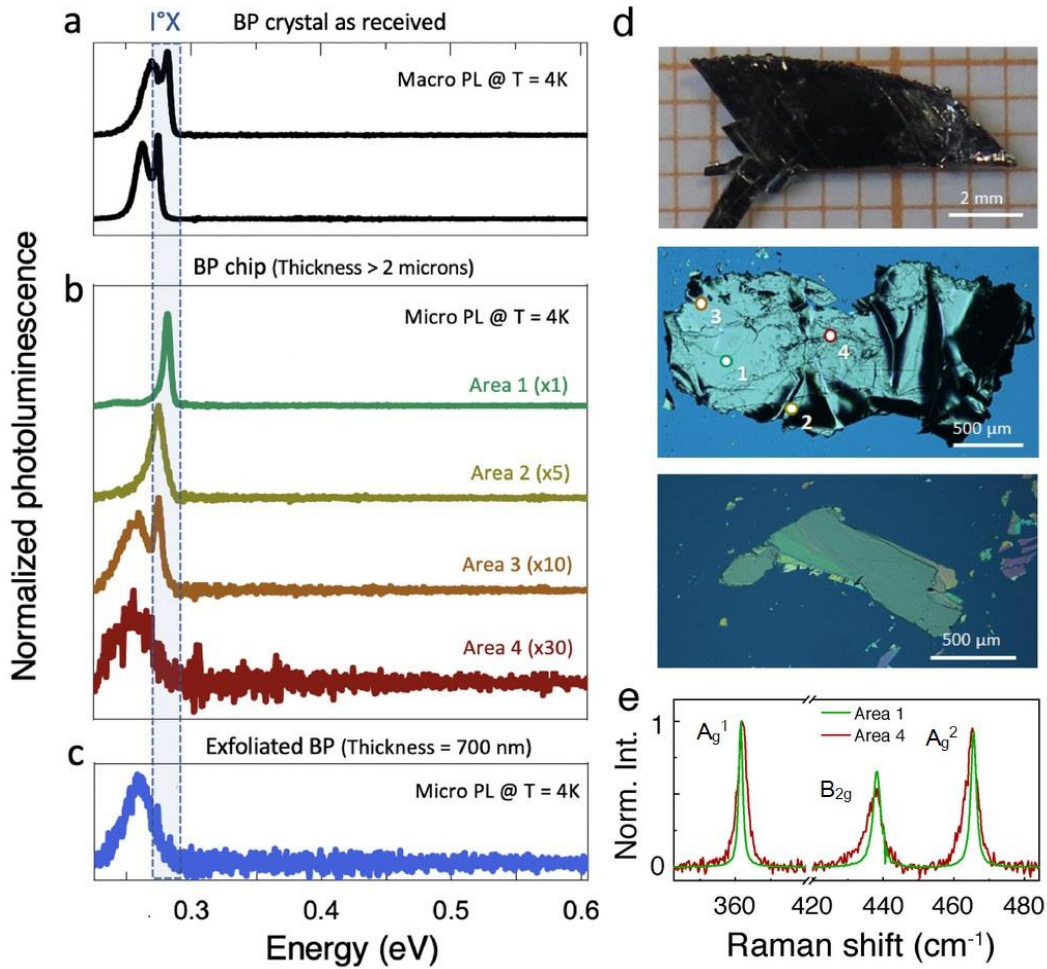

**Figure 1**. Photoluminescence spectra recorded a 4 K from **(a)** an as-received BP crystal, **(b)** from a thick BP chip obtained after a single-stamping exfoliation step, and **(c)** from a 700 nm thick flake obtained after multiple-stamping exfoliation steps. The excitation wavelength is 532 nm and the fluence is $2.10^2$ W.cm$^{-2}$ for the macro-PL (a) and $3.10^3$ W.cm$^{-2}$ for the micro-PL (b,c) experiments. **(d)** Photography and optical images corresponding to the three samples analysed in (a,b,c) from top to bottom. Labels 1 to 4 refer to the areas where spectra in (b) are recorded. **(e)** Raman spectra recorded from areas 1 and 4 of the sample (b) with a 633 nm excitation laser.

To investigate it, we operated a single-stamping exfoliation of a BP crystal, i.e. we stamped the crystal with PDMS only once, hence obtaining a BP chip of micron scale thickness, corresponding to an intermediate thickness sample of a measure between the size of the BP crystal and the ultrathin samples commonly reported in literature. Low temperature micro-PL spectra recorded at different selected areas of this BP chip (areas labeled 1 to 4 in the image of the sample in Figure 1d) interestingly exhibit different PL line shapes (Figure 1b). On one hand, a bright and narrow band dominates the emission fingerprint for areas 1 and 2. The emission energy and linewidth present strong similarities with those of the I°X bound exciton found in the pristine crystal[21]. Furthermore, micro-Raman measurements performed in the same region indicate a good structural quality with a narrow linewidth for all BP Raman modes, (see Figure 1e). We associated these spectral signatures to a low-defective zone of the BP sample where the radiative pathway is driven by bound excitons. On the other hand, the area 4 displays a broad emission band centered at lower energy, similar to the low-energy emission band of the BP crystal reported by Carré et al.[21]. We observed that this area is characterized by an overall broadening of the Raman modes (Figure 1e), suggesting the presence of an additional defect density caused by the exfoliation process. An intermediate situation with the coexistence of the two bands is observed in some regions of the chip, alike area 3. They reproduce the overall macro PL spectrum recorded on the crystal. Our micro-PL and micro-Raman analysis of BP samples resulting from single- and multiple-stamping exfoliations clearly indicate that the emergence of this band is concomitant with the formation of defects upon exfoliation.

In the light of these observations, we conclude that the exfoliation based on PDMS stamping, commonly used for producing BP samples, induces deep changes in the nature of the luminescent process in BP layers, resulting in a broad emission band that may rely to the low-energy band observed in the pristine crystal in macro PL. The anisotropic structure of BP and the low value of the Young modulus, compared to other 2D materials, could explain the high sensitivity of BP to the mechanical exfoliation processing.

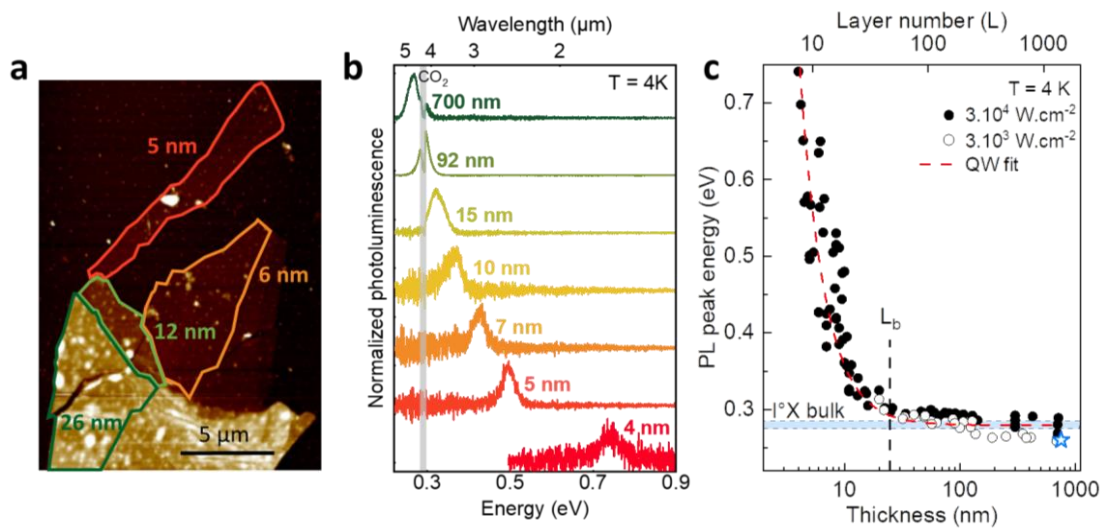

**Figure 2** (**a**) Representative AFM image of exfoliated BP and deposited on Si/SiO$_2$. (**b**) Selection of photoluminescence spectra of BP flakes at different thicknesses probed at 4 K and excited at a wavelength of 532 nm and at a fluence of $3.10^4$ W.cm$^{-2}$. The grey line indicates the part of the spectra disturbed by the CO$_2$ absorption band. (**c**) Evolution of the PL peak energy as a function of the thickness (black bullets and circles) and its fit based on a quantum well (QW) model. (red dashed line). The bound exciton energy of the BP crystal at 4 K is marked as a horizontal blue line. A black dashed line indicates the threshold thickness at 25 nm. The blue star corresponds to the energy of the PL peak shown in Figure 1c.

**Quantum-confinement on a large thickness range**

To further investigate the characteristics of this emission peak, we probed the photoluminescence of 79 BP exfoliated flakes at 4K, with a thickness ranging from 4 nm to

700 nm (see SupMat1 in the supplementary information file for details on PL measurements procedure). The AFM image of a representative thin sample is displayed in Figure 2a whereas a set of optical and AFM images of other flakes are presented in SupMat2. Figure 2b presents the normalized PL spectra at cryogenic temperature of typical BP flakes. The grey line indicates the part of the spectra disturbed by the $CO_2$ absorption band coming from the free-space part of the collection beam in our apparatus. In all exfoliated samples we probed, the luminescence spectra consist in a single and broadened emission band. This observation is consistent with the PL spectra reported in the literature for different thicknesses[7,17,19,20] with a line width ranging from 15 meV to 45 meV. Note that the PL band from the thickest flakes coincides with the broadened band reported in Figure 1c. For each flake, we associate the energy of the emission to the maximum of the PL spectrum. When it falls in the $CO_2$ absorption range (gray rectangle in Figure 2b), for example for the 92 nm film, the PL signal is modeled with a Gaussian peak and the energy at maximum is extrapolated.

The evolution of the PL peak energy over the wide thickness interval considered, from about 10 to about 1000 layers, follows two distinct regimes with a transition thickness ($L_b$) around 25 nm (about 50 layers), as highlighted by a black dashed line in Figure 2c.

At thicknesses higher than $L_b$, the PL energy evolves moderately, from 0.27 eV to 0.32 eV, and remains close to that of the bound exciton (0.275 eV at 2 K, green line in Figure 2c). These values are, in average, slightly lower than those presented in the literature for the same thickness range (0.32 eV[20] and 0.31 eV[17] for 70 nm and 220 nm-thick BP layers probed at 80K, respectively) but this can be explained by the fact that the gap of black phosphorus blueshifts with temperature[22,23]. Measures have been taken at two different excitation fluences: black bullets and empty circles correspond to laser powers of $3 \times 10^4$ and $3 \times 10^3$ W·cm$^{-2}$, respectively. No degradation was observed under exposure, i.e. no trace of the laser spot under the microscope and no decrease of the luminescence signal over time. We only notice a weak dependence of the peak energy on the excitation fluence, which could be ascribed to the Burstein Moss effect[24,25] (more details in supmat3).

At thicknesses lower than $L_b$, the peak energy increases significantly, reaching 0.7 eV for the 4 nm-thick BP flake, which is much higher than what is reported in literature[7] because of the different experimental conditions. The evolution of the experimental peak energies as a function of the thickness is well described by the law $E = \frac{\hbar^2 \pi^2}{2m^* L^2} + E_0$, where $E_0$ is an offset energy and $L$ is the thickness of the slab. This is the energy behavior of two independent particles of opposite sign moving freely with an effective reduced mass $m^*$ inside an infinite-barrier quantum-well (see red fit in Figure 2c and SupMat6). Fitting the law with parameters $E_0 = 0.279$ eV (Average energy of PL measurements on the 700 nm thick flake) and the measured thicknesses, we extract an effective mass of m*=0.049m$_0$ (m$_0$: mass of the electron in vacuum). Intriguingly, the effective mass extracted from our measurements is comparable to the exciton effective mass ($m_w$) predicted for the bulk by previous calculations[21,26,27] including the anisotropic Wannier model introduced by some of us[21].

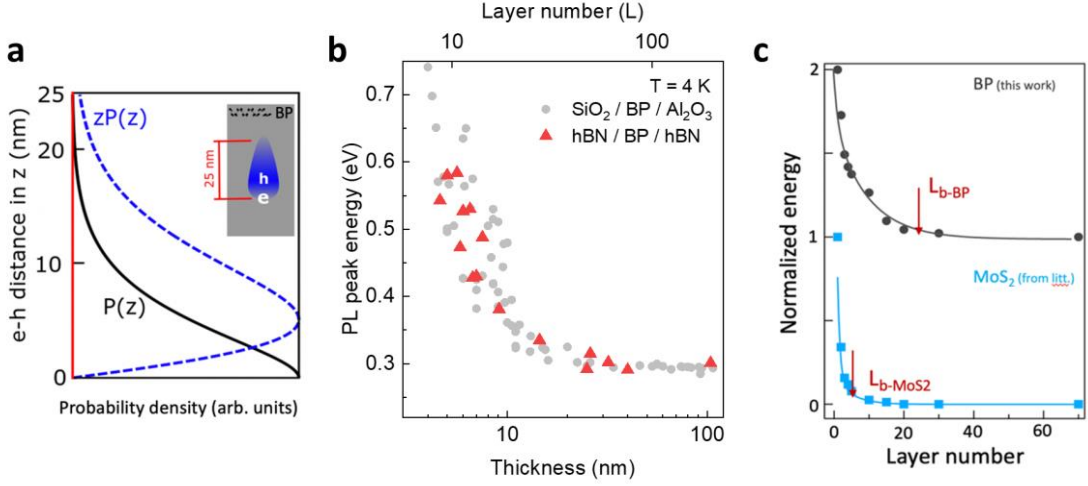

**Figure 3**. (**a**) Probability densities P(z) (black solid line) and z P(z) (blue dashed line). Inset: cartoon representing the electron-hole z-distance probability zP(z) inside a bulk-BP slab. (**b**) Variation of the photoluminescence peak energy with the thickness. hBN / BP / hBN (red triangles) and $SiO_2$/BP/$Al_2O_3$ (gray dots) heterostructures. **c** Comparison of the emission band energy variation in BP and $MoS_2$[6,28,29]. Red arrows indicate the limit beyond which the energy band gap varies by less than 3% of the total band gap modulation from the monolayer to the bulk.

We explored further this similarity by reverting to our bulk Wannier model and calculating the corresponding probability P(z) of finding an electron-hole pair of vertical extension z and the distribution of the average electron-hole pair vertical distance |z|P(z). A plot of these two quantities is reported in Figure 3a while the details of the calculation are discussed in the SupMat5. The result is that both distributions start being negligible at around $L_w = 25$ nm which indicates that, in the bulk, excitons extend in the stacking direction for $L_w = 25$ nm at most. In a BP slab, the same exciton would start "feeling'' the presence of the surfaces for thicknesses $L < L_w$, thus entering into a quantum well regime. After the similarity of the effective masses, it is remarkable that also the threshold thickness for the quantum well regime measured experimentally ($L_b$) is so close to a corresponding quantity predicted for the bulk exciton ($L_w$).

Finally, we notice that in our bulk excitonic model, the kinetic contribution to the energy was much more sensitive to changes of the parameters than the potential contribution. If this is still the case in the BP slabs, then the PL signature should not be affected by changes of the dielectric environment. Experimentally, we tested this hypothesis by comparing the evolution of the photoluminescence energy of some hBN / BP / hBN heterostructures (hBN has a dielectric constant of about 3[30]) with the $SiO_2$/BP/$Al_2O_3$ samples studied above (the dielectric constant of alumina is around 9[31]). More details on the heterostructures fabrication and their luminescence are given in the SupMat4. Results are reported in Figure 3b in red triangles (hBN encapsulated) and grey bullets (alumina-passivated samples of Figure 2). The superposition of the PL energy as a function of thickness demonstrate that, indeed, the quantum confined emission is not significantly affected by changes of the dielectric environment, at least in this range of thicknesses. This can be explained by the fact that, at these thicknesses, the wavefunctions of the particles are much more strongly affected by the confinement than the screening of their interaction. We notice also that this feature justifies "a posteriori" the independent-particle approximation implicit in the two-particle quantum well model, which nicely describes the observed emission behavior.

The characteristics of this quantum well regime depict a quite unique situation among the 2D materials. Indeed it is found to start below a threshold thickness $L_b$ of about 25 nm, corresponding to about 50 layers and the increase of the peak energy is quite slow upon reducing thickness as attested by the very low effective mass of particle pair ($m^*=0.049m_0$) involved in the model. For a comparison, in $MoS_2$ the PL energy enters the quantum well regime at a thickness of about 5 layers ($L_b$=2-3 nm) and increases steeply (see Figure 3c). This slow change of the emission energy may be an advantage for possible applications because it makes the modulation of the emission easier to control. Furthermore, we observed that these characteristics are robust against changes of the substrate or the passivation layer, which may be another advantage as it simplifies the integration of BP slabs in actual devices.

Finally, we noticed that these two quantities (the threshold thickness and the pair effective mass) together with the low sensitivity to changes in the dielectric medium are common also to the bulk excitonic model we introduced in a previous study. These similarities suggest that excitons are somehow involved in the formation and the evolution of the broadened emission band we studied.

**Conclusions**

To conclude, our work reports the evolution of luminescence at cryogenic temperature of many exfoliated slabs of black phosphorus over a wide and so far unexplored range of thicknesses.

Without being able to identify clearly the origin of the broad luminescence emission peak from exfoliated BP samples, our measures and analysis allows us to formulate the hypothesis that the robust quantum well emission we observed comes from excitons whose recombination pathways are altered by some structural disorder or defects introduced by the exfoliation procedure itself. This peak coincides with the low-energy band of the bulk crystal spectrum at very large thickness. As a consequence, this emission has properties that are unique to black phosphorus slabs. They are related to the high softness of the material, which explains the impact of the exfoliation on the photoluminescence properties, and to the low effective mass of paired particles that explains the exceptionally wide range of thicknesses over which the quantum well emission can be observed, contrasting with other semiconducting 2D materials.

**Acknowledgments.**

The authors acknowledge funding from the French national research agency (ANR) under the grant agreement No. ANR-17-CE24-0023-01 (EPOS-BP). This project has also received funding from the European Union's Horizon 2020 research and innovation program under grand agreement N° 881603 (Graphene Flagship core 3).

**References:**

1. Cheng, J., Wang, C., Zou, X. & Liao, L. Recent Advances in Optoelectronic Devices Based on 2D Materials and Their Heterostructures. *Advanced Optical Materials* **7**, 1800441 (2019).

2. Han, W. Perspectives for spintronics in 2D materials. *APL Materials* **4**, 032401 (2016).

3. Xia, F., Wang, H. & Jia, Y. Rediscovering black phosphorus as an anisotropic layered material for optoelectronics and electronics. *Nat Commun* **5**, 4458 (2014).

4. Li, L. *et al.* Black phosphorus field-effect transistors. *Nature Nanotech* **9**, 372–377 (2014).

5. Tran, V., Soklaski, R., Liang, Y. & Yang, L. Layer-controlled band gap and anisotropic excitons in few-layer black phosphorus. *Phys. Rev. B* **89**, 235319 (2014).

6. Mak, K. F., Lee, C., Hone, J., Shan, J. & Heinz, T. F. Atomically Thin MoS 2 : A New Direct-Gap Semiconductor. *Phys. Rev. Lett.* **105**, 136805 (2010).

7. Chen, C. *et al.* Bright Mid-Infrared Photoluminescence from Thin-Film Black Phosphorus. *Nano Lett.* **19**, 1488–1493 (2019).

8. Favron, A. *et al.* Photooxidation and quantum confinement effects in exfoliated black phosphorus. *Nature Mater* **14**, 826–832 (2015).

9. Island, J. O., Steele, G. A., van der Zant, H. S. J. & Castellanos-Gomez, A. Environmental instability of few-layer black phosphorus. *2D Mater.* **2**, 011002 (2015).

10. Qiao, J., Kong, X., Hu, Z.-X., Yang, F. & Ji, W. High-mobility transport anisotropy and linear dichroism in few-layer black phosphorus. *Nat Commun* **5**, 4475 (2014).

11. Gaufrès, E. *et al.* Momentum-Resolved Dielectric Response of Free-Standing Mono-, Bi-, and Trilayer Black Phosphorus. *Nano Lett.* **19**, 8303–8310 (2019).

12. Li, L. *et al.* Direct observation of the layer-dependent electronic structure in phosphorene. *Nature Nanotech* **12**, 21–25 (2017).

13. Pei, J. *et al.* Producing air-stable monolayers of phosphorene and their defect engineering. *Nat Commun* **7**, 10450 (2016).

14. Yang, J. *et al.* Optical tuning of exciton and trion emissions in monolayer phosphorene. *Light Sci Appl* **4**, e312–e312 (2015).

15. Zhang, S. *et al.* Extraordinary Photoluminescence and Strong Temperature/Angle-Dependent Raman Responses in Few-Layer Phosphorene. *ACS Nano* **8**, 9590–9596 (2014).

16. Wang, X. *et al.* Highly anisotropic and robust excitons in monolayer black phosphorus. *Nature Nanotech* **10**, 517–521 (2015).

17. Zhang, Y. *et al.* Wavelength-Tunable Mid-Infrared Lasing from Black Phosphorus Nanosheets. *Adv. Mater.* **32**, 1808319 (2020).

18. Ye, L. *et al.* Highly polarization sensitive infrared photodetector based on black phosphorus-on-WSe 2 photogate vertical heterostructure. *Nano Energy* **37**, 53–60 (2017).

19. Chen, C. *et al.* Widely tunable mid-infrared light emission in thin-film black phosphorus. *Sci. Adv.* **6**, eaay6134 (2020).

20. Wang, J. *et al.* Mid-infrared Polarized Emission from Black Phosphorus Light-Emitting Diodes. *Nano Lett.* **20**, 3651–3655 (2020).

21. Carré, E. *et al.* Excitons in bulk black phosphorus evidenced by photoluminescence at low temperature. *2D Mater.* **8**, 021001 (2021).

22. Villegas, C. E. P., Rocha, A. R. & Marini, A. Anomalous temperature dependence of the band-gap in Black Phosphorus. *Nano Lett.* **16**, 5095–5101 (2016).

23. Baba, M., Nakamura, Y., Shibata, K. & Morita, A. Photoconduction of Black Phosphorus in the Infrared Region. *Jpn. J. Appl. Phys.* **30**, L1178–L1181 (1991).

24. Burstein, E. Anomalous Optical Absorption Limit in InSb. *Phys. Rev.* **93**, 632–633 (1954).

25. Moss, T. S. The Interpretation of the Properties of Indium Antimonide. *Proc. Phys. Soc. B* **67**, 775–782 (1954).

26. Asahina, H., Shindo, K. & Morita, A. Electronic Structure of Black Phosphorus in Self-Consistent Pseudopotential Approach. *J. Phys. Soc. Jpn.* **51**, 1193–1199 (1982).

27. Takao, Y. & Morita, A. Electronic structure of black phosphorus: Tight binding approach. *Physica B+C* **105**, 93–98 (1981).

28. Scheuschner, N. *et al.* Photoluminescence of freestanding single- and few-layer $MoS_2$. *Phys. Rev. B* **89**, 125406 (2014).

29. Splendiani, A. *et al.* Emerging Photoluminescence in Monolayer $MoS_2$. *Nano Lett.* **10**, 1271–1275 (2010).

30. Laturia, A., Van de Put, M. L. & Vandenberghe, W. G. Dielectric properties of hexagonal boron nitride and transition metal dichalcogenides: from monolayer to bulk. *npj 2D Mater Appl* **2**, 6 (2018).

31. Vila, R., González, M., Mollá, J. & Ibarra, A. Dielectric spectroscopy of alumina ceramics over a wide frequency range. *Journal of Nuclear Materials* **253**, 141–148 (1998).

# Quantum well confinement and competitive radiative pathways in the luminescence of black phosphorus layers

**Authors**

Etienne Carré[1,2], Lorenzo Sponza[1], Alain Lusson[2], Ingrid Stenger[2], Sébastien Roux[1,2], Victor Zatko[3], Bruno Dlubak[3], Pierre Seneor[3], Etienne Gaufrès[4], Annick Loiseau[1] and Julien Barjon[2]*

[1] Laboratoire d'Étude des Microstructures (LEM), UMR 104 CNRS-Onera, Université Paris Saclay, 92322 Châtillon, France.
[2] Groupe d'Etude de la Matière Condensée (GEMaC)), CNRS-UVSQ, Université Paris Saclay, 45 Avenue des Etats-Unis 78035 Versailles, France
[3] Unité Mixte de Physique, CNRS, Thales, Université Paris-Saclay, 91767 Palaiseau, France.
[4] Laboratoire Photonique Numérique et Nanosciences, CNRS - Institut d'Optique - Univ. Bordeaux, 33400 Talence, France

# 1. Experimental methods

**Exfoliation and passivation of BP thin layers:**

Thin layers of black phosphorus are exfoliated from a high quality crystal (HQ Graphene – 99.995% purity) in glove box under argon atmosphere (<0.5 ppm $O_2$, <1 ppm $H_2O$) to prevent its photo-oxidation under ambient conditions[1]. Conventional PDMS (1:10 ; cross-linked 1h at 80° C) buffers are used to reduce BP single crystals to thin layers, which are then deposited on a $SiO_2$(300 nm)/Si substrate. For each sample, the area of interest is identified with an optical microscope and the thickness is measured with an Atomic Force Microscope (Bruker Innova AFM). These measurements are done within the glove box. A hermetically argon filled suitcase was used to transport and introduce BP samples in an atomic layer deposition (ALD) growth chamber (BENEQ TFS 200) for passivation purposes. A 10 nm continuous $Al_2O_3$ has been deposited on BP samples from the reaction of tri-methyl-aluminum and $H_2O$, which provides an efficient barrier against air exposure[2,3]. From the storage of the crystal to the passivation of the thin layers, the BP is never in contact with ambient air and oxidizing agents, which results in good quality samples.

**Luminescence experiments:**

Micro photoluminescence measurements are performed at 4 K with a He cold finger cryostat. The excitation source is a frequency-doubled Nd:YAG Laser (532 nm). It is focused on the BP flakes with a laser spot of about 2 µm diameter thanks to a reflective objective lens (Cassegrain Thorlabs, x25, NA 0.4). The PL signal is collected back through the objective and detected with a Fourier Transform Infra-Red (FTIR) spectrometer (BOMEM DA8, $CaF_2$ or Quartz beamsplitters, spectral resolution 0.5 meV) equipped with an InSb detector cooled at 80 K. A chopper and a lock-in amplifier are needed to eliminate the undesirable blackbody radiations in the infrared spectral range.

Macro photoluminescence measurements are quite similar. The set-up differs in two ways: the cryostat used is a helium bath and the laser is no longer focused by an objective but by a lens allowing it to reach a diameter of about 100 µm.

**RAMAN experiments:**

The Raman spectrum was recorded in backscattering configuration using a high-resolution Raman spectrometer setup (Labram HR800 from HORIBA Jobin-Yvon). A He-Ne laser operating at 633 nm is focused on the sample through a 100x objective with a numerical aperture of 0.80 to for an illumination spot diameter of about 1 µm. To avoid any form of degradation, the laser power is set below 2 mW and the measurements are performed under argon flow to avoid BP oxidation.

## 2. Raman measurement

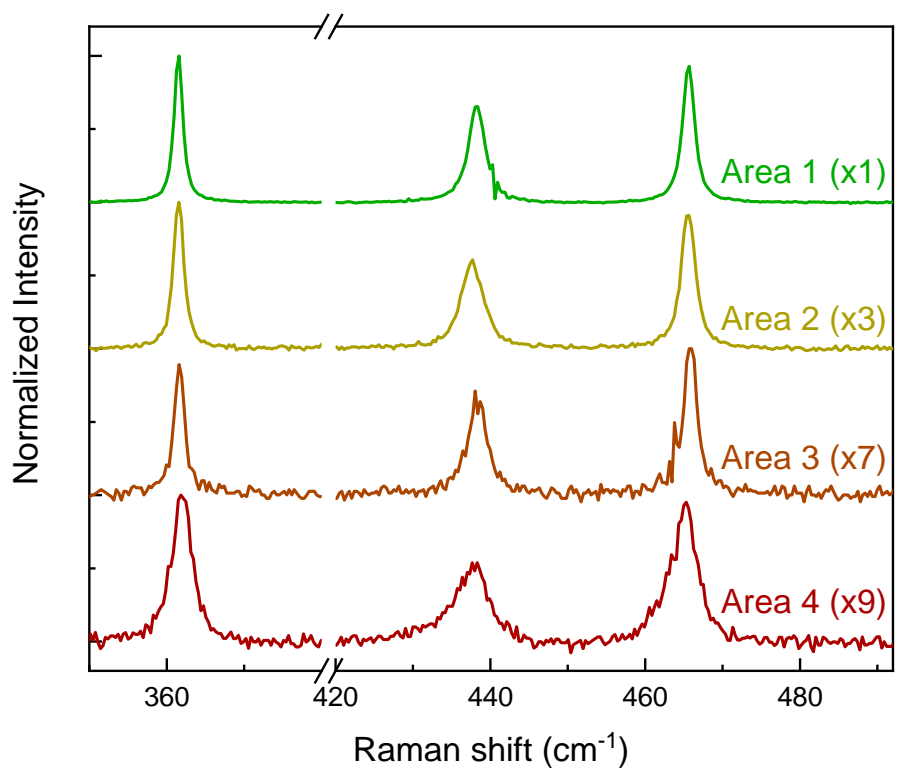

**Figure SM2** – Raman spectra recorded from areas 1 to 4 of Figure 1b sample with a 633 nm excitation laser.

# 3. AFM images

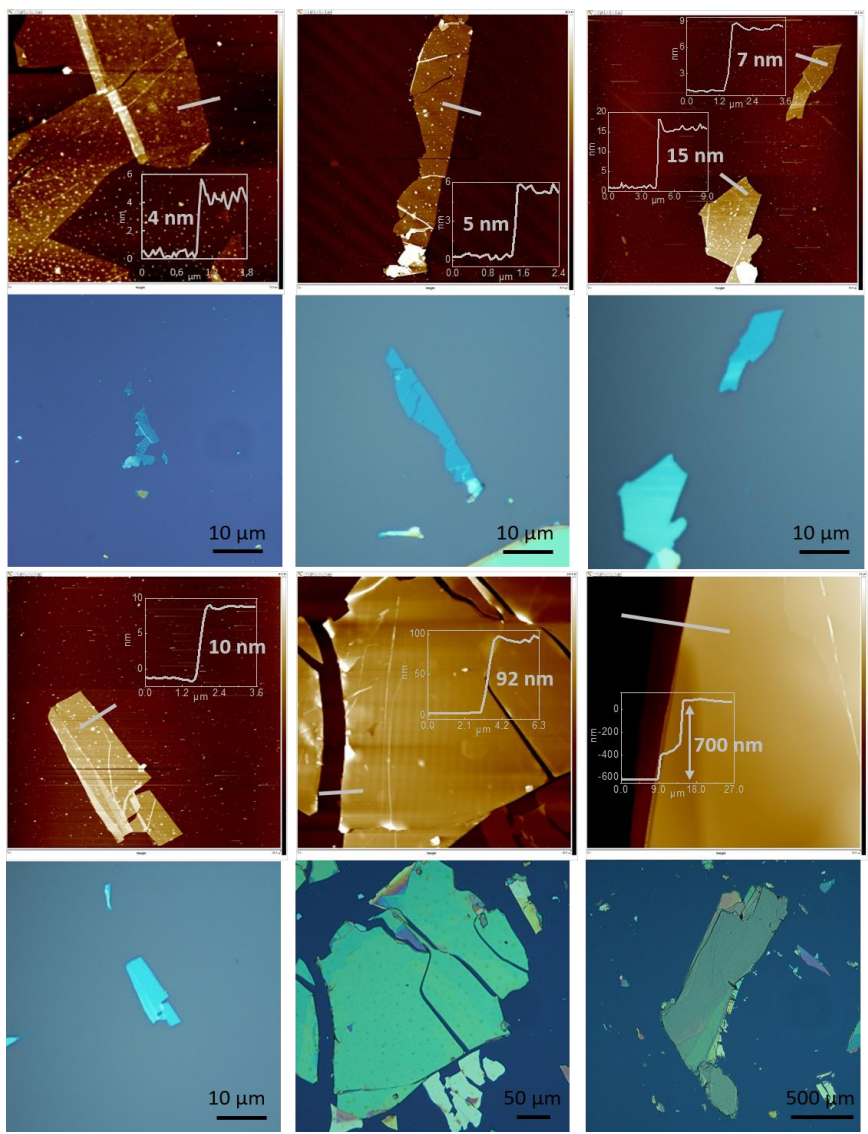

**Figure SM3** - Optical and AFM images of the seven flakes from which the spectra shown in Figure 2b are recorded. AFM profiles in insets of the images are extracted along the grey lines marked in the AFM images.

# 4. Effect of laser fluence

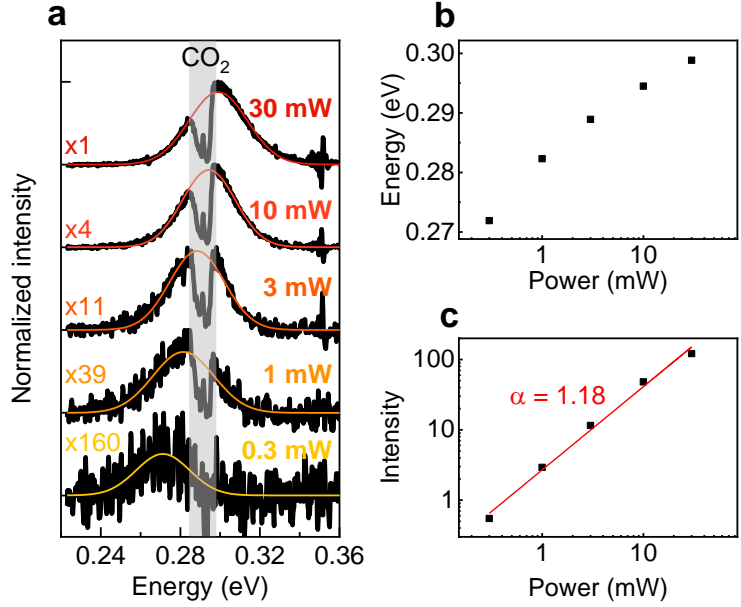

**Figure SM4 -** (a) PL spectrum power dependence at 4 K measured on a 65 nm thick BP flake. The shaded area represents the $CO_2$ absorption zone. The PL energy (b) and the integrated intensity (c) as a function of excitation power are extracted from Gaussian fits shown in color lines.

Figure SM4a shows the power dependence of a typical thick BP flake PL spectrum at 4 K. As mentioned before, we fit the PL signal with a Gaussian to extract the energy (Figure SM4b) as well as the integrated intensity (Figure SM4c) of the peak, partially in the area of the $CO_2$ absorption. It is worth noting that the PL spectrum does not widen with power and remains stable around 30 meV, which attests that the heating due to the beam seems to be negligible.

The PL spectrum shows a clear blueshift of few tens of meV over two decades of excitation power. This behavior is due to the filling of the conduction band by the high number of carriers generated by the high LASER power. This energy difference is commonly known as the Burstein-Moss shift (BMS)[4,5]. We can estimate the density of state in the conduction band of the BP at a given temperature (1) :

$$N_C = 2 \cdot \left(\frac{2\pi m_e^* kT}{h^2}\right)^{\frac{3}{2}} \qquad (1)$$

With $m^*_e = 0.413 m_0$, the effective electron mass averaged over the 3 crystallographic axes[6], we find a density of state in the conduction band at 4 K of $N_C(4\ K) = 1.024 \cdot 10^{16}$ cm$^{-3}$. A similar calculation allows us to obtain the density of state in the valence band of the BP : $N_V(4\ K) = 7.482 \cdot 10^{15}$ cm$^{-3}$.

This value must be compared with the charge carrier density created in photoluminescence by a 532 nm laser. We can estimate the diameter of the beam $d_{beam}$ (2) as well as the penetration depth $p_{depth}$ (3) from the complex refractive index N measured in the literature[7]. The excitation being circularly polarized, we take the average of the absorption coefficients following armchair and zigzag.

$$d_{beam} = 1.22 \frac{\lambda}{NA} = 1.6 \text{ μm} \quad (2)$$

$$p_{depth} = \frac{1}{\alpha} = \frac{\lambda}{4\pi.Im(N)} \sim 180 \text{ nm} \quad (3)$$

We then find an excitation volume of V = $3.7.10^{-13}$ cm³. Assuming a beam reflection of 0 and a quantum efficiency of 1 we can then calculate the density of injected charge carriers $n_{eh}$ for a 30 mW laser power (4). The lifetime of the charge carriers does not seem to be known for the BP so we choose to take an arbitrary value of τ = 0.1 ns which is the same order of magnitude as the lifetime in other classical semiconductors[8].

$$(4) \quad n_{eh} = \frac{P.\tau.\lambda}{h.c} . \frac{1}{V} = 2.2.10^{19} cm^{-3}$$

We then find that $n_{eh}$ >> $N_C$(4 K), $N_V$(4 K), therefore, the strong photo excitation regime forces us to consider the filling effects of the conduction band (CB). As the excitation power increases, more electrons are sent to the CB and thermalize at the band edge. The newly generated electrons cannot access the band edge states and are then forced to occupy higher energy states and increase the optical gap, this is what is observed in figure 2b. BMS effect is also observed on BP samples by several experimental measurements requiring similar excitation powers[9–11].

In Figure SM4c, the integrated intensity of the PL I is fitted with a classical power model $I = P^k$, with P the incident laser power and k a factor that depends on the nature of the radiative transition[12,13]. The over linear behavior of the PL peak intensity (k = 1.18) suggests an excitonic nature (radiative de-excitation of a bound or free exciton) rather than a defect-related emission, generally characterized by a sublinear k-factor. However, this analysis should be completed, in particular by extending the power range on which this study is made.

# 5. Heterostructures

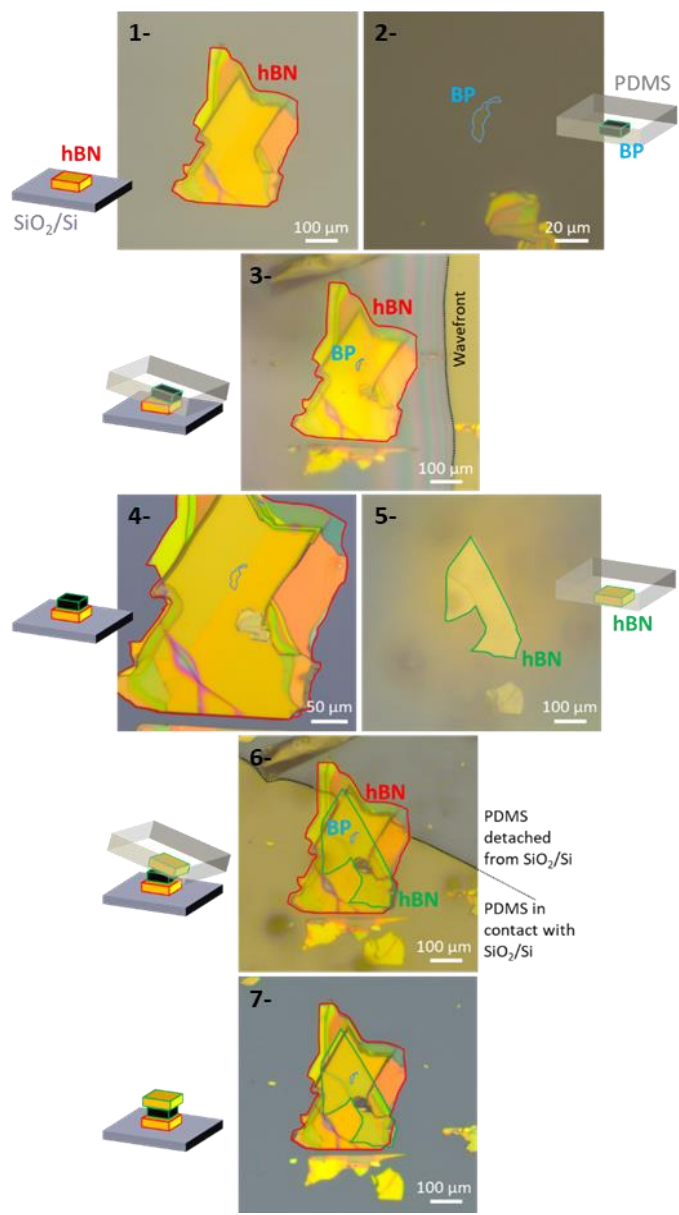

**Figure SM5a -** Optical images and associated diagrams describing the manufacturing process of a hBN/BP/hBN heterostructure: 1- Crystal of hBN on SiO$_2$/Si. 2- Crystal of BP on PDMS. 3- BP transfer on hBN, the wavefront corresponds to the PDMS - SiO$_2$/Si contact line. 4- BP/hBN heterostructure on SiO$_2$/Si. 5- hBN crystal on PDMS. 6- Transfer of hBN on BP/hBN. 7- hBN/BP/hBN heterostructure.

Figure SM5a displays the different steps necessary to the fabrication of a hBN/BP/hBN heterostructure. A high pressure / high temperature hexagonal Boron Nitride (hBN) crystal[14] is classically exfoliated using PDMS and deposited on a SiO$_2$(300 nm)/Si substrate (1-). Using the "all dry" method described by Castellanos Gomez et al.[15], an exfoliated BP flake is spotted

on PDMS by optical contrast (2-) and then slowly deposited on a large flat hBN flake (3-). Before encapsulation, the thickness of the BP flake is measured by AFM (4-). Similarly, another large flat hBN flake is spotted by optical contrast on the PDMS (5-) and deposited on the top of the pre-existing BP/hBN stack (6-) to build a hBN/BP/hBN heterostructure (7-). The entire process is carried out in a glove box under inert atmosphere to avoid BP oxidation. About 15 similar stacks were made with different black phosphorus thicknesses as shown in Figure 3a.

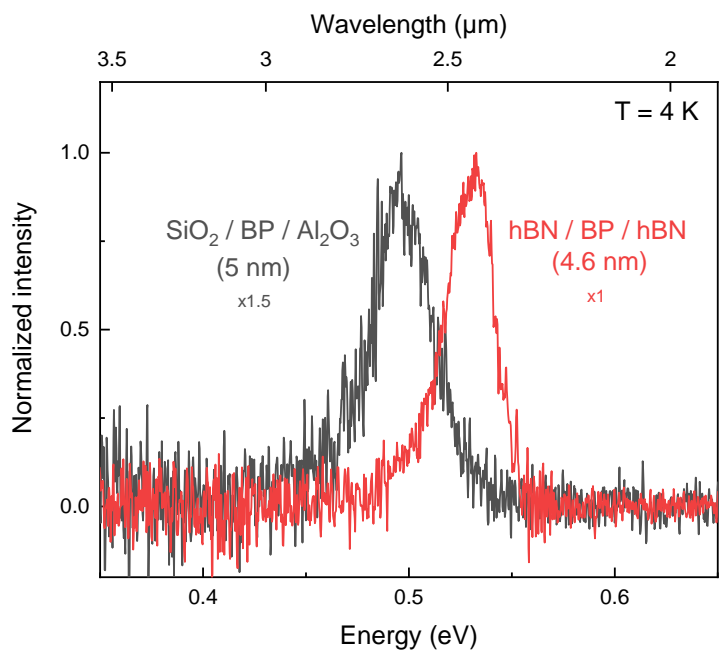

**Figure SM5b -** PL spectra of a 5 nm thick alumina-passivated black phosphorus flake (gray) and a hBN / BP (4.6 nm thick) / hBN (red) heterostructure at 4 K. PL intensities are normalized (normalization factors are x1.5 and x1 respectively as indicated in the figure).

Figure SM5b shows typical spectra of two BP flakes of comparable thickness, one encapsulated in hBN flakes (red) and the other one passivated with alumina (gray) at 4 K. The PL spectrum of BP encapsulated in hBN does not differ drastically from that of alumina passivated BP. In particular the intensities and full widths at half maximum of the peaks (29 meV for BP encapsulated in hBN ; 37 meV for passivated BP) are of the same order of magnitude contrary to what could be observed for the atomic layers of TMDC encapsulated in boron nitride[16]. The energy difference between the two peaks (0.53 eV for BP encapsulated in hBN ; 0.49 meV for passivated BP) can be explained by the slight difference in BP thickness. The variation of the dielectric environment does not seem to influence the luminescence energy at these thicknesses as it had been shown for the atomic layers of BP[17].

# 6. Two-particle quantum well

In a two-particle quantum well model, two charges (in the following labelled *e* and *h*) are free to move inside a region confined by two infinite potential barriers placed at –L/2 to L/2 (i.e. no spill-out density is allowed). The corresponding two-particle wavefunction writes

$$\phi_L(z_e, z_h) = \frac{1}{\sqrt{N}} \cos\left(\frac{\pi z_e}{L}\right) \cos\left(\frac{\pi z_h}{L}\right), \text{ with } N \text{ a normalization constant.}$$

The associated kinetic energy, also referred as confinement energy, is

$$<K> = -\frac{\hbar^2}{2} \int\int \phi_L \left\{ \frac{1}{m_e}\frac{\partial^2 \phi_L}{\partial z_e^2} + \frac{1}{m_z}\frac{\partial^2 \phi_L}{\partial z_h^2} \right\} dz_e\, dz_h = \frac{\hbar^2 \pi^2}{2m^* L^2}$$

where $m_z$ are the effective masses of the single charges and $m^* = \frac{m_e m_h}{m_e + m_h}$ is the reduced effective mass of the pair.

The fit reported in Figure 2 gives an effective mass of the same order of magnitude of the the experimental values used to predict the excitonic properties in the bulk model[18]. We also stress that this is an independent-particle model because there is no particle-particle interaction term. The adoption of such assumption to describe our measurements is justified *a posteriori* by the observation that the emission is insensitive to changes of the dielectric environment.

# References

1.  Favron, A. *et al.* Photooxidation and quantum confinement effects in exfoliated black phosphorus. *Nature Mater* **14**, 826–832 (2015).

2.  Wood, J. D. *et al.* Effective Passivation of Exfoliated Black Phosphorus Transistors against Ambient Degradation. *Nano Lett.* **14**, 6964–6970 (2014).

3.  Galceran, R. *et al.* Stabilizing ultra-thin black phosphorus with in-situ -grown 1 nm-$Al_2O_3$ barrier. *Appl. Phys. Lett.* **111**, 243101 (2017).

4.  Burstein, E. Anomalous Optical Absorption Limit in InSb. *Phys. Rev.* **93**, 632–633 (1954).

5.  Moss, T. S. The Interpretation of the Properties of Indium Antimonide. *Proc. Phys. Soc. B* **67**, 775–782 (1954).

6.  Narita, S. *et al.* Far-Infrared Cyclotron Resonance Absorptions in Black Phosphorus Single Crystals. *J. Phys. Soc. Jpn.* **52**, 3544–3553 (1983).

7.  Asahina, H., Maruyama, Y. & Morita, A. Optical reflectivity and band structure of black phosphorus. *Physica B+C* **117–118**, 419–421 (1983).

8.  Pelant, I. & Valenta, J. *Luminescence spectroscopy of semiconductors*. (Oxford University Press, 2012).

9.  Hedayat, H. *et al.* Non-equilibrium band broadening, gap renormalization and band inversion in black phosphorus. *2D Mater.* **8**, 025020 (2021).

10. Whitney, W. S. *et al.* Field Effect Optoelectronic Modulation of Quantum-Confined Carriers in Black Phosphorus. *Nano Lett.* **17**, 78–84 (2017).

11. Lin, C., Grassi, R., Low, T. & Helmy, A. S. Multilayer black phosphorus as a versatile mid-infrared electro-optic material. *Nano Lett.* **16**, 1683–1689 (2016).

12. Schmidt, T., Lischka, K. & Zulehner, W. Excitation-power dependence of the near-band-edge photoluminescence of semiconductors. *Phys. Rev. B* **45**, 8989–8994 (1992).

13. Spindler, C., Galvani, T., Wirtz, L., Rey, G. & Siebentritt, S. Excitation-intensity dependence of shallow and deep-level photoluminescence transitions in semiconductors. *Journal of Applied Physics* **126**, 175703 (2019).

14. Taniguchi, T. & Watanabe, K. Synthesis of high-purity boron nitride single crystals under high pressure by using Ba–BN solvent. *Journal of Crystal Growth* **303**, 525–529 (2007).

15. Castellanos-Gomez, A. *et al.* Deterministic transfer of two-dimensional materials by all-dry viscoelastic stamping. *2D Mater.* **1**, 011002 (2014).

16. Cadiz, F. *et al.* Excitonic Linewidth Approaching the Homogeneous Limit in $MoS_2$-Based van der Waals Heterostructures. *Phys. Rev. X* **7**, 021026 (2017).

17. Gaufrès, E. *et al.* Momentum-Resolved Dielectric Response of Free-Standing Mono-, Bi-, and Trilayer Black Phosphorus. *Nano Lett.* **19**, 8303–8310 (2019).